\documentclass{aa}
\usepackage{txfonts}
\usepackage{graphicx}
\begin{document}

 \title{The Active Quiescence of HR Del (Nova Del 1967)
 \thanks{ Based on  observations made with the International
Ultraviolet Explorer  and de-archived from the ESA  VILSPA
Database. }}
\subtitle{The Ex-Nova HR Del }

   \author{Pierluigi Selvelli
          \inst{1}
          \and Michael
 Friedjung
          \inst{2}
    }
   \offprints{P. Selvelli}

   \institute{CNR-IASF-Osservatorio Astronomico di Trieste  -  
Via
Tiepolo 11, 34131 Trieste, Italy
              \\
              \email{selvelli@ts.astro.it}
         \and
             Institut d'Astrophysique, 98 Boulevard Arago, 75014 
Paris, France\\              \email{fried@iap.fr}
             \
             }

   \date{Received .....; accepted }

  \abstract {
This new  UV study  of  the ex-nova HR Del is based on all of the
data obtained with  the  International
Ultraviolet Explorer (IUE) satellite,  and includes the important
series of spectra taken in 1988 and 1992 that have not been
analyzed so far. This has allowed us to make a detailed study of
both the long-timescale  and the short-timescale  UV variations,
after the return of the nova, around 1981-1982, to the 
pre-outburst
optical magnitude. After the correction for the reddening
($E_{B-V}=0.16$), adopting a distance $d $=850 pc we  have 
derived
a mean UV luminosity close to $L_{UV}$$\sim$ 56 $L_{\odot}$, the
highest value among classical novae in "quiescence". Also the
"average"  optical absolute magnitude ($M_{v}=+2.30$) is 
indicative
of  a  bright object.   The UV continuum  luminosity,  the  HeII
1640 \AA~  emission  line  luminosity,   and  the  optical 
absolute
magnitude all give a  mass accretion rate $\dot{M}$  very close
 to 1.4$\times$10$^{-7}$$ M_{\odot}$ yr$^{-1}$ ,  if one
assumes that the luminosity of the old nova is due to a
non-irradiated accretion disk.   The UV
continuum has declined by a factor less than 1.2 over the 13 
years
of the IUE observations, while the UV emission lines have faded 
by
larger factors. The continuum distribution is well fitted with
either a black body of 33,900 K, or a power-law
$F_{\lambda}$$\sim$$\lambda$$^{-2.20}$. A comparison with the 
grid
of models of Wade and Hubeny (1998) indicates  a low $M_{1}$ 
value
and a relatively high $\dot{M}$ but the best fittings to the
continuum and the line spectrum come from different  models.
We show that the  "quiescent" optical magnitude at
$m_{v}$$\sim$12 comes from the hot component  and not from the
companion star. Since most IUE observations correspond to the
"quiescent" magnitude at $m_{v}$$\sim$12, the same as in the
pre-eruption stage, we infer that the pre-nova, for at least 70
years prior to eruption, was also very bright at near the same
$L_{UV}$, $M_{v}$, $\dot{M}$, and T values as derived in the
present study for the ex-nova. The  wind components in the P Cyg
profiles of the  CIV  1550 \AA~ and NV 1240 \AA~ resonance lines
are strong and variable on short timescales,  with  $v_{edge} $ 
up
to -5000 km s$^{-1}$, a remarkably high value. The phenomenology 
in
the short-time variations of the wind indicates the presence of 
an
inhomogeneous  outflow.
We discuss the nature of the strong UV
continuum and wind features and the implications of the presence 
of
a "bright" state a long time before and after outburst on our
present knowledge of the pre-nova and post-nova behavior.
 \keywords{ Stars: novae  -  Ultraviolet: stars   - Stars: winds 
}
 }

\maketitle
%

\section{Introduction}

HR Del=Nova Del 1967   brightened in July 1967  to a magnitude
of  5.5 (Alcock 1967) from a pre-nova magnitude near
$m_{v}$$\sim$12 (Stephenson 1967; Barnes \& Evans 1970; Robinson
1975). The  object remained for
as long as 5 months near this pre-maximum halt and brightened 
again
in mid Dec. 1967 to reach a maximum peak of  $m_{v}$$\sim$3.5
(Terzan 1970; Terzan et al. 1974;  Bartolini et al.
1969; Mannery 1970). Both the initial rise of only  7 magnitudes
 and
the very long timescale to reach maximum are rather unusual for a
nova. The decline after maximum was  irregular (in May 1968 the
brightness increased again to $m_{v}$ 4.3) and extremely slow 
with
$t_{3}$ about 225 days  (Rafanelli \& Rosino 1978). The beginning
of the nebular phase occurred about one year after outburst and  
the return to the pre-outburst visual magnitude occurred after
1975 (Drechsel et al. 1977; Bruch 1982; Rafanelli \& Rosino
1978) or even after 1981-1982 to
judge from the lightcurve of the AAVSO
 and from  the IUE FES counts  (see Section 4).
At the present time the  visual magnitude of  HR Del
shows small oscillations around $m_{v}$=12.0.

\begin{figure*}
\centering
\includegraphics{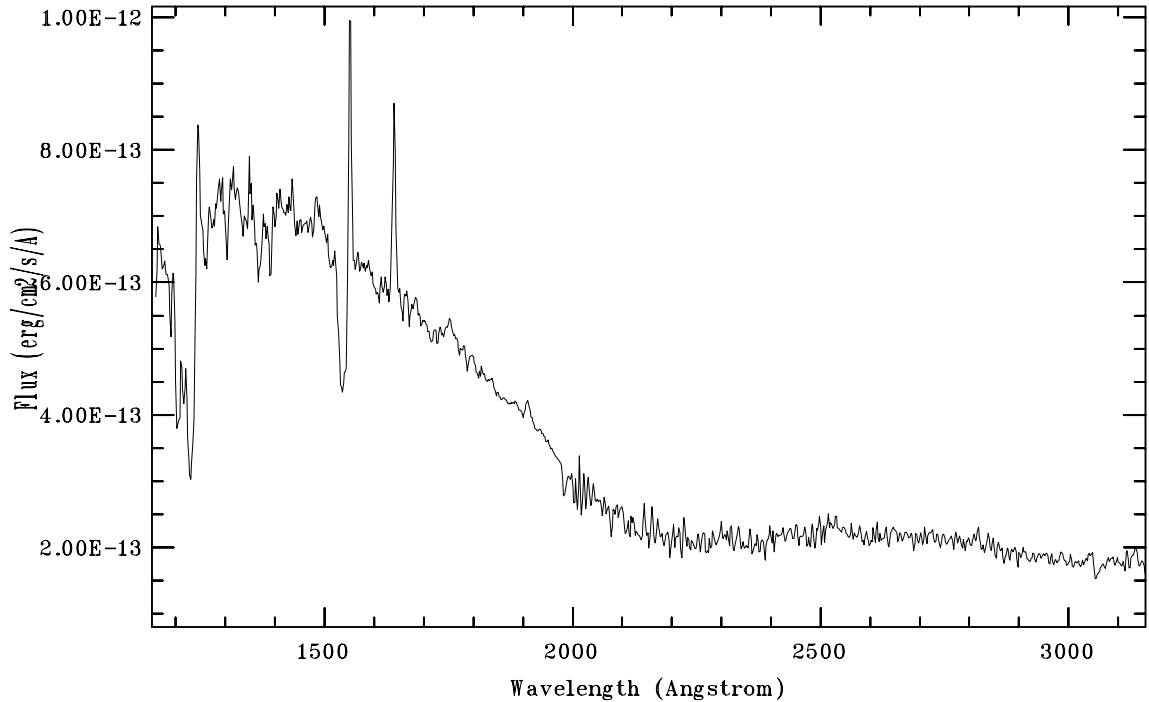}
\caption{The "average" 1979-1988 UV spectrum. This "virtual"
spectrum has been obtained by averaging and merging all  of the
SW and LW  IUE spectra obtained from 1979 to 1988. 
Outstanding features are the P Cyg profiles in the  CIV
1550 \AA~ and NV 1240 \AA~ resonance lines. The shortward 
displaced
 absorption component of this latter line is blended with an
interstellar Ly$_{\alpha}$ line. Other spectral features are the
HeII 1640 \AA~ emission line and the absorption lines of SiII 
1260
\AA~ and CII 1335 \AA.}
\end{figure*}

\begin{figure*}
\centering
\includegraphics{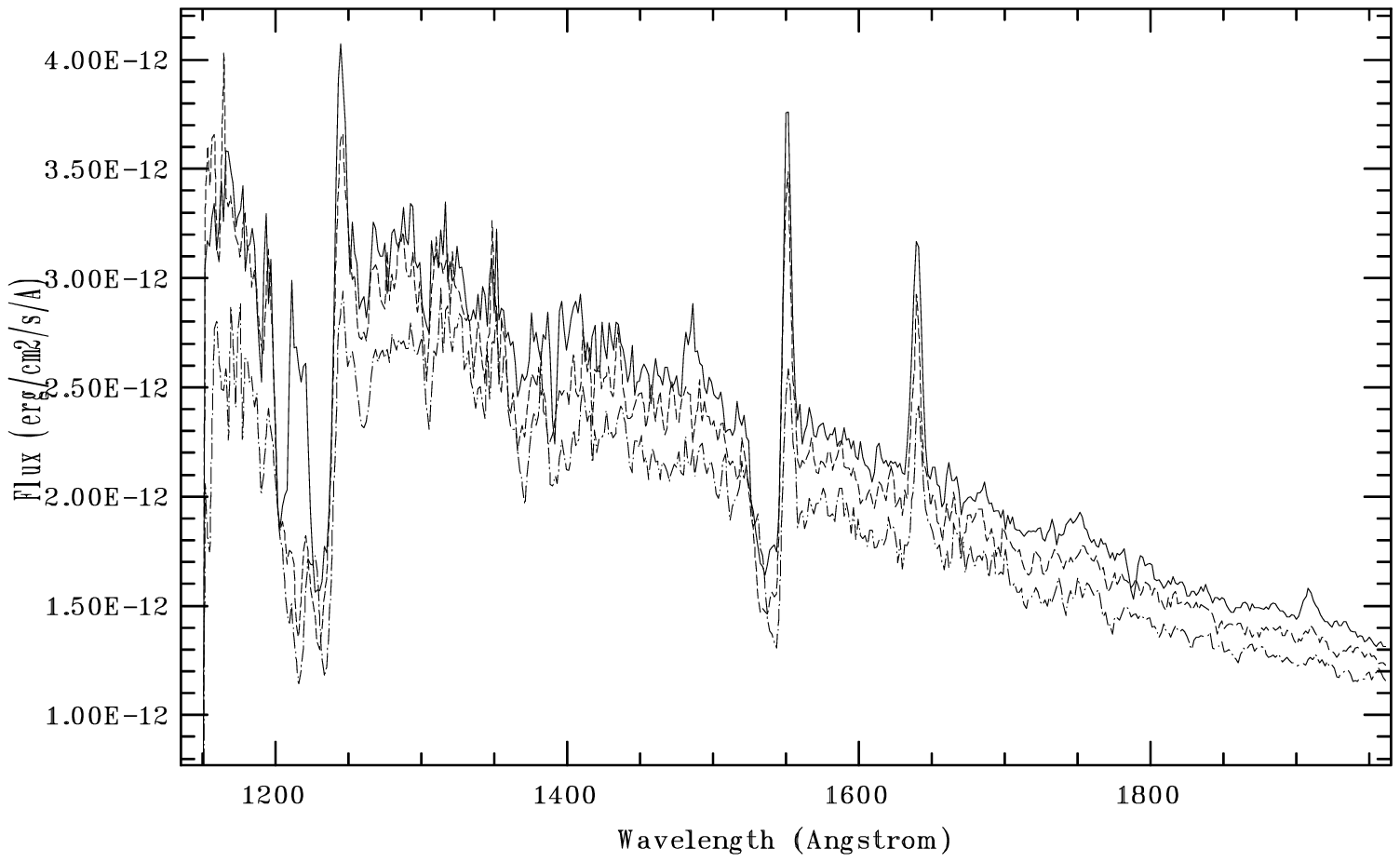}
\caption{The  long-timescale  variations in the SWP range.
The  "average" SWP spectra  of  1980 (top, continuous line), 1988
(mid, dash line), and 1992
(bottom, dot-dash  line) are displayed on an absolute scale.
  The spectra  are corrected for reddening  with  
E$_{B-V}$=0.16.}.
\end{figure*}

Spectroscopic observations made
during the first outburst phases revealed expansion velocities  
in
the range from -200 to -700 km s$^{-1}$ (Hutchings 1968)  but
values of -1200 km s$^{-1}$  and up  to -1800 km s$^{-1}$ have
been reported during the late  decline phases in 1968 
(Wallerstein
1968;  Rafanelli \& Rosino 1978).  Friedjung (1992), from a study
of the pre-maximum spectral development has pointed out the 
unusual
nature of nova HR Del, and suggested that HR Del might only
marginally satisfy the conditions for a thermonuclear runaway.

\begin{figure*}
\centering
\includegraphics{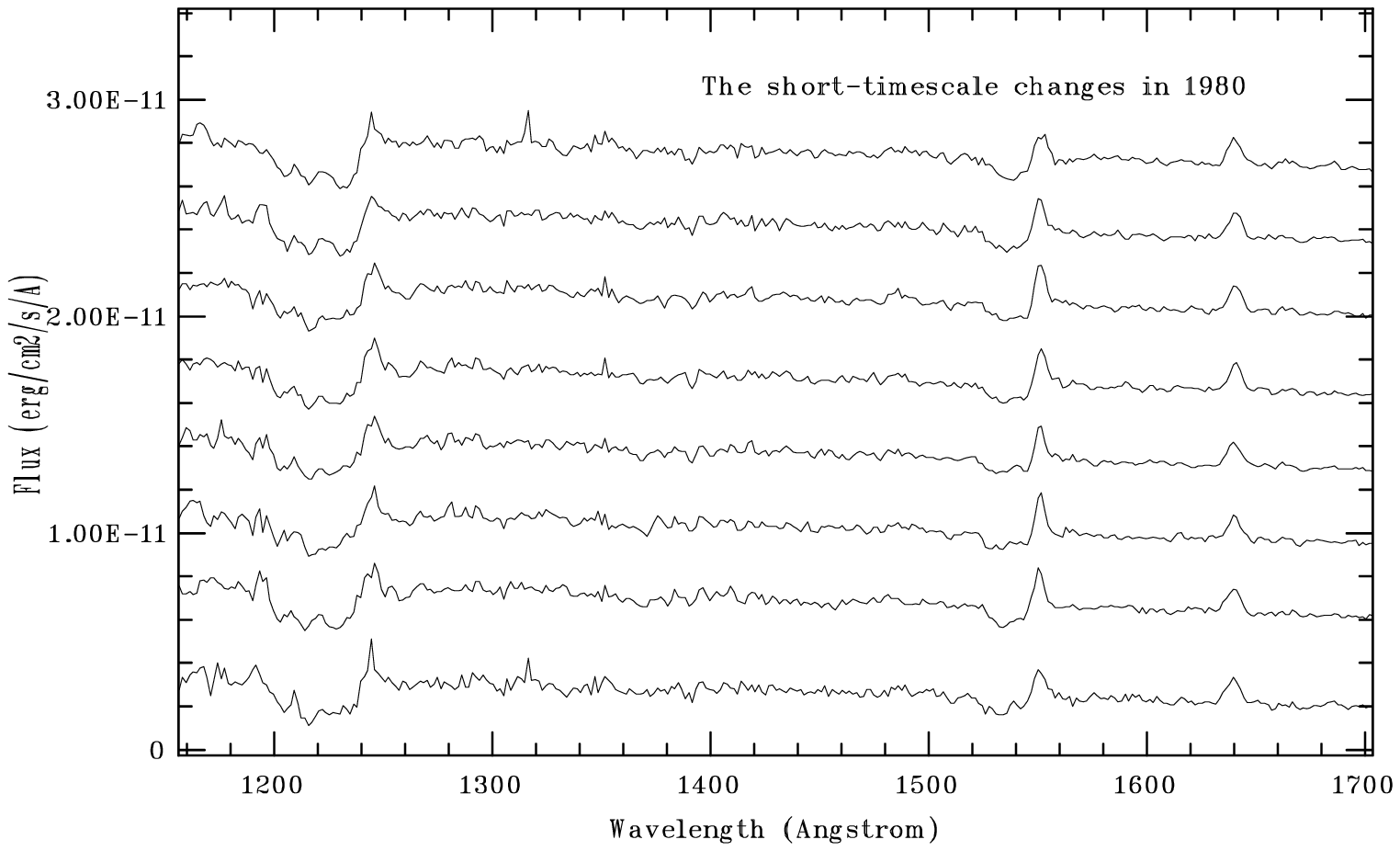}
\caption{Display of the short-timescale variations in the spectra
of the sequence  of Aug. 21,  1980.
 The spectra have been shifted  vertically  for
display purposes. The spectra  are corrected for  reddening  with
E$_{B-V}$=0.16. The  average time separation between successive
exposures is near 49 min.}
\end{figure*}

\begin{figure*}
\centering
\includegraphics{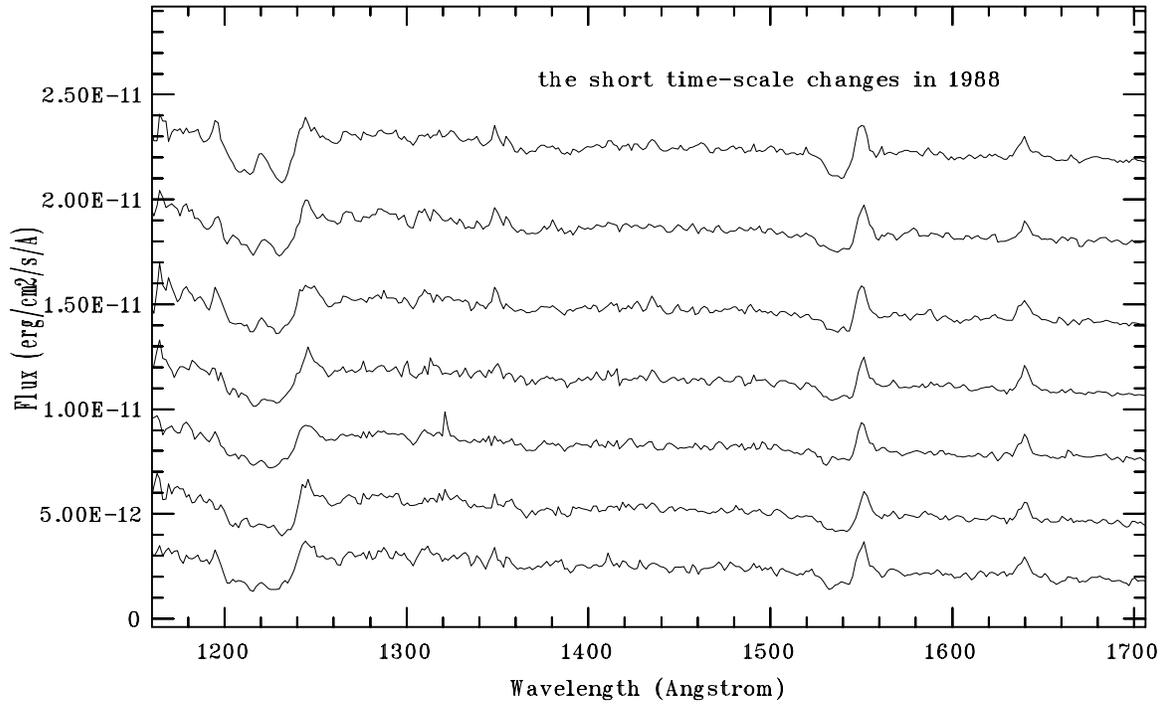}
\caption{The same as in Fig.3 for the sequence of Apr. 29, 1988.
The average time separation is near 59 min.  }
\end{figure*}

\begin{figure*}
\centering
\includegraphics{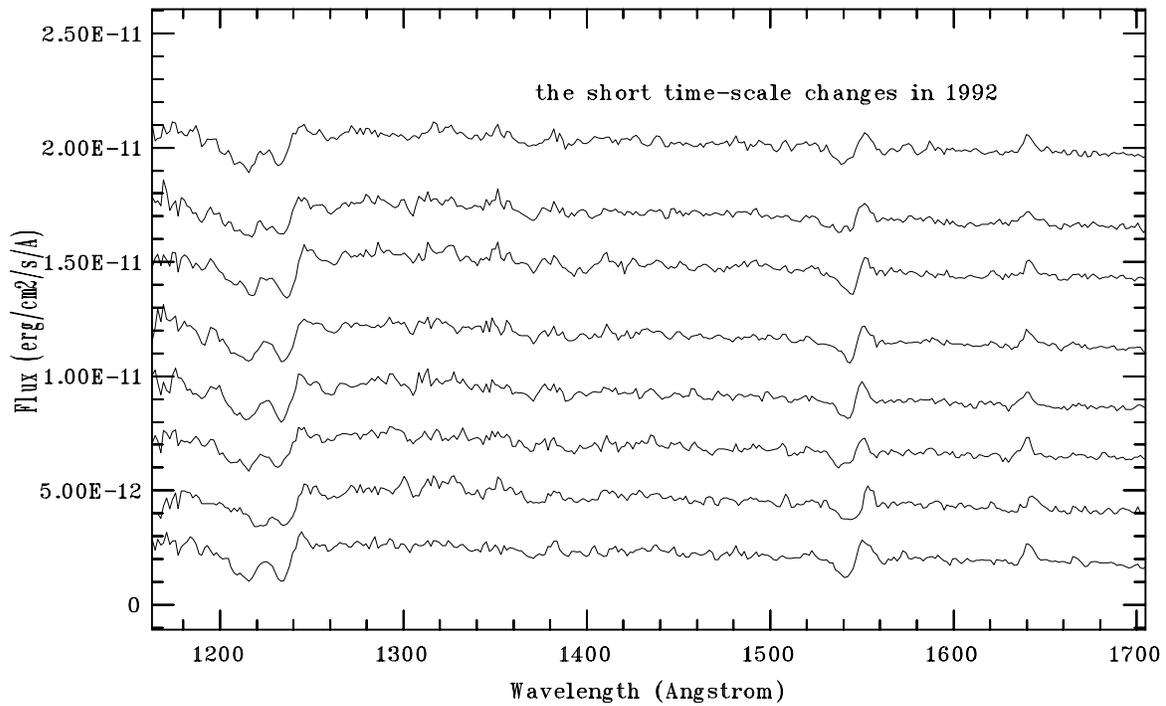}
\caption{The same as  in Fig.3 for the sequence of   Aug. 30, 
1992.
 The average time separation is near 52 min.}
\end{figure*}

\begin{figure*}
\centering
\includegraphics{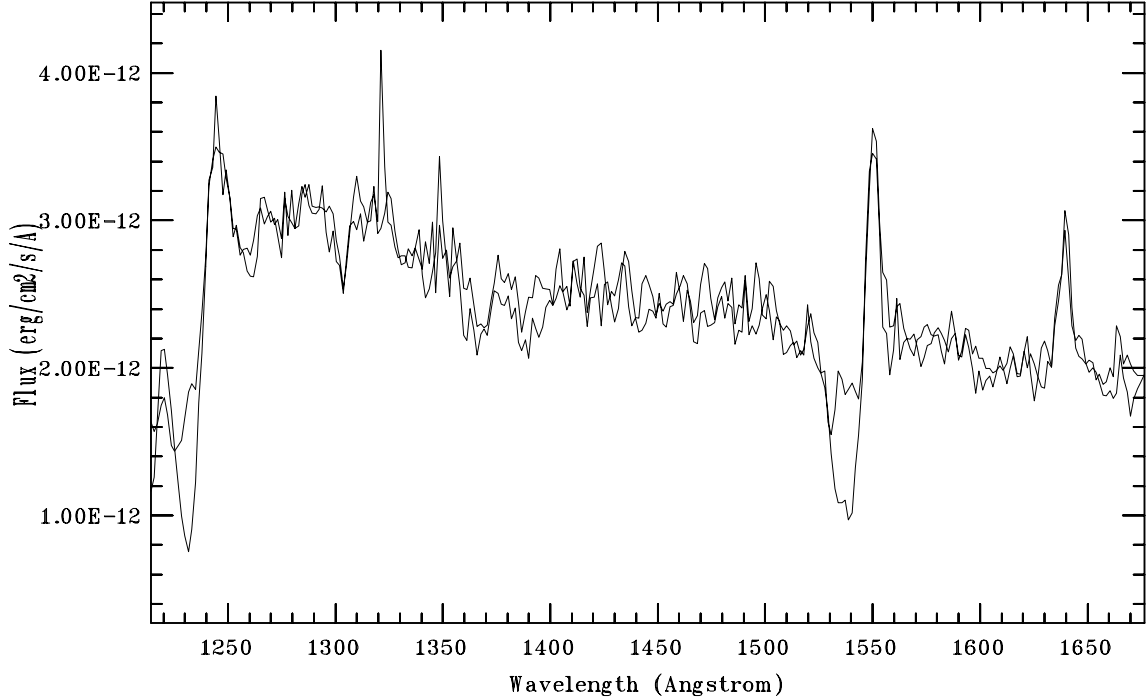}
\caption{The  dramatic short-timescale  changes in the 
absorption components of the CIV 1550  \AA ~ and NV 1240 \AA
~lines  in two spectra   (SWP33398 and SWP33402)
of the sequence of  April 29, 1988. The changes
are remarkably similar  in  both lines  The  time separation
between the start of the
two exposures  is of $3^h$$53^m$. } 
 \end{figure*}

\begin{figure*}
\centering
\includegraphics{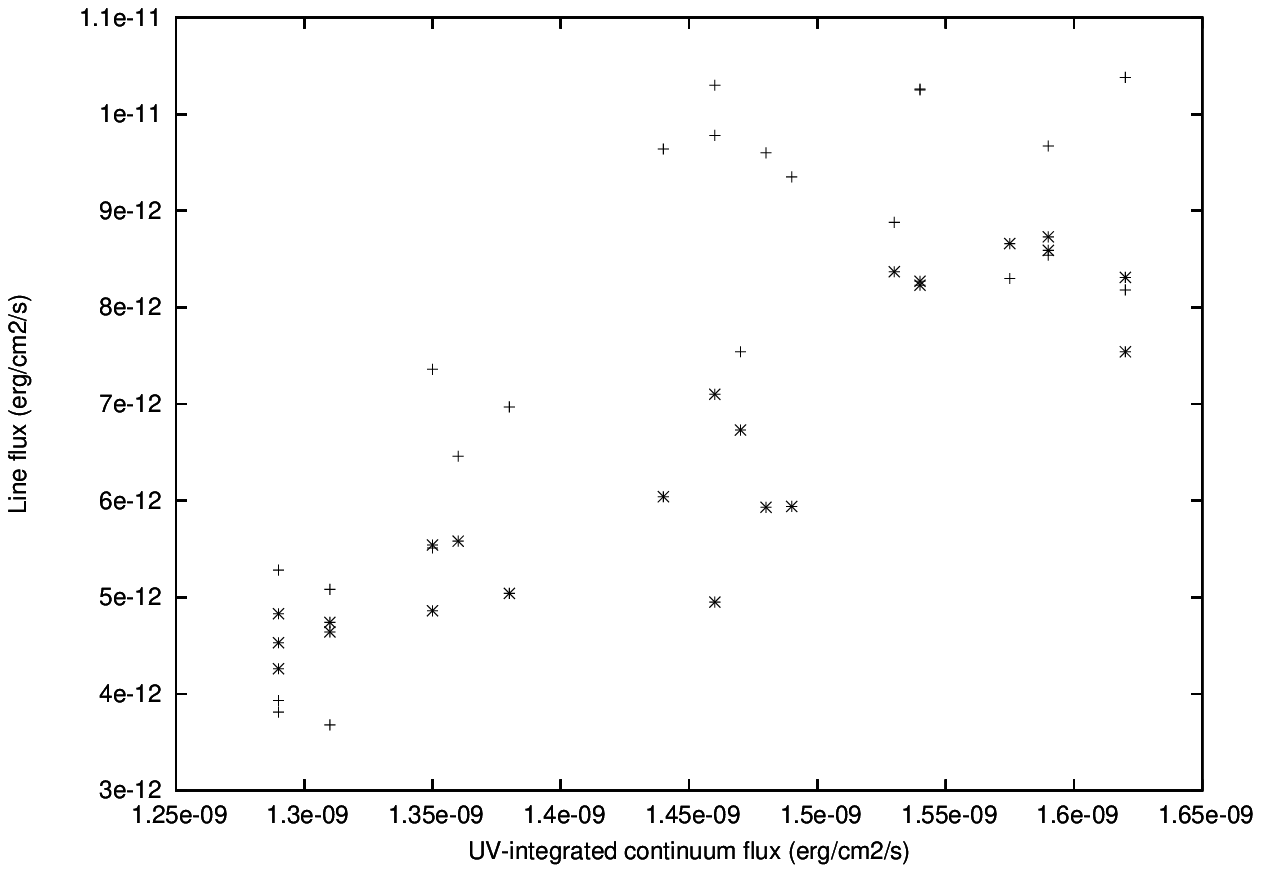}
\caption{The overall correlation between   the
far-UV-integrated  continuum flux and the
line flux  in the  CIV 1550 \AA ~ (plus symbols)
and HeII 1640 \AA~(asterisks   symbols)  emission  lines.}
 \end{figure*}

The ejected shell was first observed by Kohoutek (1981) and 
Solf (1983). Recent observations (Slavin et al.
1994; Slavin, et al.  1995) have revealed  a
structure  with two polar caps and an
equatorial ring  but there are some discrepancies  in the exact
shell size  (see also Downes \& Duerbeck 2000).   Very
recently,  O' Brien  et al. (2002)  have  reported  v=-290 km
s$^{-1}$ and
v=-580 km s$^{-1}$  for the expansion  speed in the equatorial 
ring
and in the  polar  caps   respectively  and derived a distance d
$\sim 1100\pm100 $pc.
The position angle of the elongated remnant, as found in
different studies, is close to 45$^{o}$.

The distance, as estimated by several authors using
various techniques (expansion parallax, interstellar lines,
MMRD, etc.) is  near 900$\pm$200 pc. The orbital period of
the system is  $0^{d}.214165$ and the orbital  inclination is
close to 40$^{o}$  (Bruch 1982, Kuerster \& Barwig 1988).

Being one of the brightest  nova remnants, HR Del was the target 
of
several IUE observations made in 1979-1980,
1988, and 1992,  but the  UV literature is based only on
the spectra  taken in 1979-80 (Krautter et al. 1981;
Rosino et al. 1982; Friedjung et al. 1982).
These studies all  agree upon the presence of a hot continuum and
that of strong P Cyg profiles in the CIV 1550  \AA~lines, but,
not surprisingly, a quite wide range of continuum temperatures 
and
outflow velocities have been derived even for the same sets of
spectra.

We have undertaken this new analysis of all IUE-INES  spectra
of HR Del with the purpose of  fully exploiting  the content  of
the IUE-INES databank by investigating both the long timescale
variation (over more than a decade) and the short timescale
variations (over a few hours ) in its UV spectrum.

\section {The IUE data }

From the IUE-INES archive  we have  retrieved both the resampled
image (SILO)  and the extracted spectrum
(MXLO) for the whole set of 49 spectra  of HR Del.  For a
description of the IUE-INES system see Rodriguez-Pascual et al.
(1999). Inspection of the SILO images has revealed a good
centering in all spectra, and the absence of  geo-coronal
Ly$_{\alpha}$ emission, except for a few spectra around SWP07108.
All SILO images  have been carefully examined for the presence of
spurious emission features, blemishes, etc. In the full set of 49
spectra, four SWP and two LWR spectra were taken with the small
aperture and are of limited use as far as absolute quantities 
such
as continuum and line emission intensities are involved. Also, 
one
spectrum (SWP05757) is badly overexposed in most of it and has 
been
excluded from the sample. Of the remaining 42 spectra (31 SWP + 
11
LW ) the most interesting ones are the 23 SWP spectra belonging 
to
the three series of spectra of  Aug. 21, 1980, Apr. 29, 1988 and
Aug. 30, 1992  since, being
obtained in a close sequence over a time baseline slightly longer
than one orbital period,  they  allow a  detailed study of the
short-timescale  variations.

We recall that the IUE data extraction and  calibration methods
have undergone several revisions during and after the IUE  
lifetime
and that this has resulted in
 non negligible changes both in the quality of the line spectrum
and
in shape of the continuum curve (see Gonzalez-Riestra et al. 
2001).

\section { The spectrum and the reddening }

The individual spectra are quite similar to each other in the
continuum and line features.
This  justifies  the creation of a "virtual"
"average" spectrum by  co-adding and merging all SW and LW 
spectra
for the epochs (1979, 1980, and 1988) in  which data for  both
spectral ranges are available (Fig. 1). In 1992 only SWP data 
were
taken and have not been included in this "average".
The improved S/N in the "average" spectrum has allowed  both the
detection of weak line features and an accurate determination of
the reddening : $E_{B-V}=0.16 \pm 0.02 $, as estimated by
applying  the common method of removing  the  2175 \AA ~bump.  In
the present study  we will adopt
A$_{v}$=0.315$\times$0.16$\sim$0.50.

Outstanding  spectral features  in most of the UV spectra of  HR
Del are the  strong P Cyg profiles in the CIV 1550 \AA ~   and
 NV 1240 \AA ~ resonance lines,  together with the
HeII   1640 \AA ~   (pure) emission line.
The  emission component  in the  P Cyg profile  of   the
NV 1240 \AA ~   line   has significantly  faded
in  the most  recent spectra   (1992).
Nebular  lines (i.e.,  NIV
1483 \AA, OIII 1666 \AA, NIII  1750 \AA,  and CIII 1909 \AA) are
clearly present only in the spectra taken in 1979-1980. The
absorption features that are present below $\lambda$  1400  \AA~
are identified as SiII 1190 \AA,  SiII 1260 \AA, OI+SiII 1303 
\AA,
CII 1335 \AA,  all these lines being indicative of the
spectrum  of a  B2-B7 star (cf. Rountree \& Sonneborn, 1993),
and possibly  OV 1370 \AA,  from a higher ionization stage.

An interstellar contribution to the zero-volt component in some 
of
these lines cannot be excluded   but the presence of
intrinsic variations in individual spectra (especially for the
SiII 1260 \AA ~line) and the absence of any absorption near
MgII  2800 \AA~  indicate that it is quite negligible.
An examination of the individual spectra does not reveal
substantial changes from spectrum to spectrum  with the exception
of the short-timescale  variations  in the  CIV and NV absorption
components that are described  in Section 5.
The LW region is almost featureless.

\section{ The long-timescale variations  }

Fig. 2 is a plot (on an absolute  scale) of the  average  SWP
spectra for the 1980, 1988 and 1992 epochs that indicates
an almost "gray" decay with time in the continuum
together with a  more pronounced decline  in the emission line
intensities, especially NV 1240  \AA~, CIV 1550 \AA~ and HeII 
1640 
 \AA.
The short wavelength continuum (SWP region) has declined by a
factor $\sim$1.19 from 1979-1980 to 1992 (about  1.08 from 1980
to 1988) but the emission  line decline is definitely larger:
NV 1240 \AA~ is down by $\sim$6, CIV  1550 \AA~ by $\sim$2.3
and He II 1640 \AA~  by $\sim$1.6. All nebular
lines i.e. NIV 1484 \AA, NIII 1750 \AA, and CIII 1909 \AA ~that
were clearly present in the 1980 spectra, have declined strongly
afterwards.

The pronounced  fading in the high ionization emission
lines, especially NV 1240 \AA, together with the moderate decline
of both the UV continuum and the optical magnitude over
the 13 years covered by the IUE observations would suggest
a decline  of  a very hot, compact source, likely to be
associated with  post-outburst phenomena,  combined with
recombination  in the nebula.

It is worth mentioning that at the epochs of the first set of IUE
observations  the V mag was at about  11.9  and declined to 12.0
around 1982.  Since then the star has remained at $m_{v}$
$\sim$12.0, despite  the (small) decline in the ultraviolet
region.  V optical magnitudes simultaneous with the IUE
observations have been obtained from the counts of the FES 
on-board
the IUE satellite, using the calibrations of Perez (1991) and
Fireman \& Imhoff (1989).

\section{The short-timescale  variations}

The fortunate circumstance that for the 1980, 1988 and
1992 epochs several SWP spectra were taken in a strict sequence 
of
exposures,  each sequence covering  more than one orbital
period (P$^{orb}$=$5^{h}$$6^{m}$$24^{s}$,  Bruch 1982),
it enabled us to study the short-timescale spectral variations 
and
their possible correlation with the orbital  phase.  Figures  3, 
4,
and 5 are display plots, on a convenient scale, of the individual
spectra taken in the three  sequences of  1980, 1988, and 1992.

The short timescale  variations do not show any convincing
relation  with the orbital  phase.  In a few cases the
fluctuations in the observed quantities  appear as semi-regular 
but
in most cases they appear as irregular and are probably  
associated
with transient phenomena. Irregular variations  in
the CIV 1550 \AA~ absorption component  occur with
timescales of the order of the time separation between successive
exposures ($\sim$50 min.),  but  it is likely  that this is
just  an upper limit set by the observations.
Fig. 6  is an example of the considerable variations  in the
absorption component  of the CIV 1550 \AA ~ doublet.  The
two spectra (SWP33398 and SWP33402) belong to the
1988 sequence  and were  taken with time separation of
3$^h$53$^m$ between the start of the two exposures..
 Similar variations
 have been observed also
in spectra of 1980 and 1992, although  with different time
separations.
The similar trend and pattern in  the  changes in the  NV and CIV
absorption components  indicates that the two  line forming
regions are close or somehow associated.

We  have looked for possible overall correlations between
 quantities such as the emission intensities of the CIV  1550
\AA~ and HeII  1640 \AA~emission  lines, the equivalent width of
the  absorption component  of  the  CIV
1550  \AA~ line, the far-UV integrated continuum flux,
etc.  A  definite linear  trend  is present  between the 
intensity
of the far-UV-integrated  continuum and the  emission intensity 
of
the  HeII 1640 \AA~ line, while  there is evidence of a
non-linear correlation  between  the   intensity of the
 far-UV-integrated   continuum  and  that of  the emission
intensity of the  CIV 1550 \AA~  line   (see  Fig.7).

It is  remarkable  that  the intensities of the  emission
and absorption components in the P Cyg profile of the CIV  1550
\AA~ line are  very poorly correlated.

\section {Discussion}

\subsection {The distance to HR Del and the system parameters  }

An estimate of the basic system parameters such as
the distance $d$, $M_{1}$, and  $R_{1}$ is  necessary before any
attempt at determining  $L_{UV}$, $M_{v}$, $\dot{M}$, etc.

We recall that various methods  have yielded a
distance to HR Del in the range 700-1100 pc. As an independent
check,  we add here one more estimate based on the elementary
assumption that for this extra slow nova $L$$\sim$$L_{Edd}$ near
the outburst maximum. If $M_{1}$=0.65$M_{\odot}$ (see the
following) then $L_{Edd}$=8.45$\cdot$10$^{37}$\rm  erg cm$^{-2}$
s$^{-1}$ and $M_{bol}^{max}$=-6.10.
Near maximum light novae radiate mostly in the optical and the
bolometric correction $BC$ is quite small and close to  -0.20.  
We
can  therefore  confidently assume  $M_{v}^{max}$$\sim$-5.9. The
visual magnitude at maximum $m_{v}^{max}$ is not clearly
defined but is in the range 3.5-5.0. Probably HR Del was slightly
under-Eddington in the extended pre-maximum plateau phase at
$m_{v}$=5.0, while was slightly super-Eddington (Seitter 1990) in
the sharp peak at  $m_{v}$=3.5 of December 1967. In this 
connection
let us note that according to Friedjung (1992) HR Del had 
probably
an optically thin wind before the sharp peak and that a
 super-Eddington
luminosity is not required to accelerate such a wind by radiation
pressure. If we assume as $m_{v}^{max}$ the intermediate value at
$m_{v}$=4.25, together with A$_{v}$=0.5,  we actually obtain  $d$
=850 pc, thus confirming the previous estimates based on other
methods. The MMRD relation of Della Valle \& Livio (1995) gives
$M_{v}^{max}$=-6.86  and  $d$=1180 pc if we take 
$m_{v}^{max}$=4.0.
The same relation as redetermined in the zero-point and amplitude
by Downes \& Duerbeck (2000) gives instead $M_{v}^{max}$=-6.42 
and
$d$
=960 pc. The MMRD of Downes \& Duerbeck (2000) gives
$M_{v}^{max}$=-5.99 and $d$=887 pc. It must be pointed out that
there
is  some degree of uncertainty on the correct value for $t_{2}$
that could affect all these estimates. In the present  study  we
will adopt  a rather conservative value for the distance,  that 
is
$d$=850 pc.

$M_{1}$ has been estimated from  the mass function
\begin{displaymath}\frac{(M_2\cdot sin
i)^3}{(M_1+M_2)^2}=1.037\times 10^{-7}\cdot K_1^3\cdot
P\end{displaymath}
adopting $P$=0$^{d}$.214165,  $K_{1}$=110 km s$^{-1}$
(Kuerster \& Barwig 1988), $M_{2}$=0.53$\pm$0.03$M_{\odot}$
(as the average of recent Mass-Period relationships for CVs
reported by Warner 1995; Patterson 1998; Clemens et al. 1998;
and Smith \& Dhillon 1998), and i$\sim 40^{o}$  (as given by
Kuerster \& Barwig (1988), reported in Ritter \& Kolb (1998) and
supported by the studies on the structure of the nebular shell, 
as
mentioned in  Sect.1). We have allowed a limited variation in 
both
$M_{2}$ and $i$, but, if we constrain $M_{1}$ to values larger 
than
0.5$M_{\odot}$, then the solutions are within a small interval: 
for
an  inclination  $i$  lower than $\sim$40$^{o}$  $M_{1}$ becomes
smaller than $M_{2}$, and an $M_{2}$ value as large as 0.60
$M_{\odot}$ is required, which seems at the limit of the
acceptable $M_{2}$ range.  On the other hand an inclination much
larger than i$\sim 40^{o}$ seems unlikely from the studies on the
nova shell. Also, the presence of absorption lines in the 
spectrum,
the high $L_{UV}$, and  the very high outflow velocity (see
Sect.6.7) all point to a medium-low inclination.

With the physical constraints  given above, we obtain
that the most likely values are: $M_{2}$=$0.53\pm 0.03$ 
$M_{\odot}$
and $M_{1}$ in the range  0.55 $M_{\odot}$-0.75$M_{\odot}$
for  $i$ in the range $40^{o}$-$45^{o}$.
In the following we will assume $M_{1}$=0.65 $\pm$
 0.10$M_{\odot}$.

We recall that the mean white dwarf (WD) mass $M_{1}$ in 
classical
nova systems, as estimated by nova frequency, is $\sim$ 1.1
$M_{\odot}$ (Ritter et al. 1991), while that estimated from the
observations (Smith \& Dhillon 1998) is $0.85\pm 0.05 M_{\odot}$.
Thus, the white dwarf in HR Del seems
undermassive.
Two pieces of circumstantial evidence, that is, the  very slow
speed class during outburst ($t_3$ $\sim 230^{\rm d}$) and the 
fact
that the ejected shell was quite massive, about $10^{-4} 
M_{\odot}$
( Anderson \& Gallagher 1977) are in agreement with the
theoretical expectations for an outburst on a low-mass WD  (Livio
1992). If $M_{1}$$\sim$0.65$M_{\odot}$ we obtain R$_{1}$=0.0125
$R_{\odot}$, as the  average from various  M-R
relations  in white dwarfs as reported
by  Nauenberg (1972),  Paczynski (cfr. Anderson 1988),  Eggleton
(cfr. Politano  et al. 1990),  that do not differ much from each
other near M=0.65$M_{\odot}$.

\begin{figure*}
\centering
\includegraphics{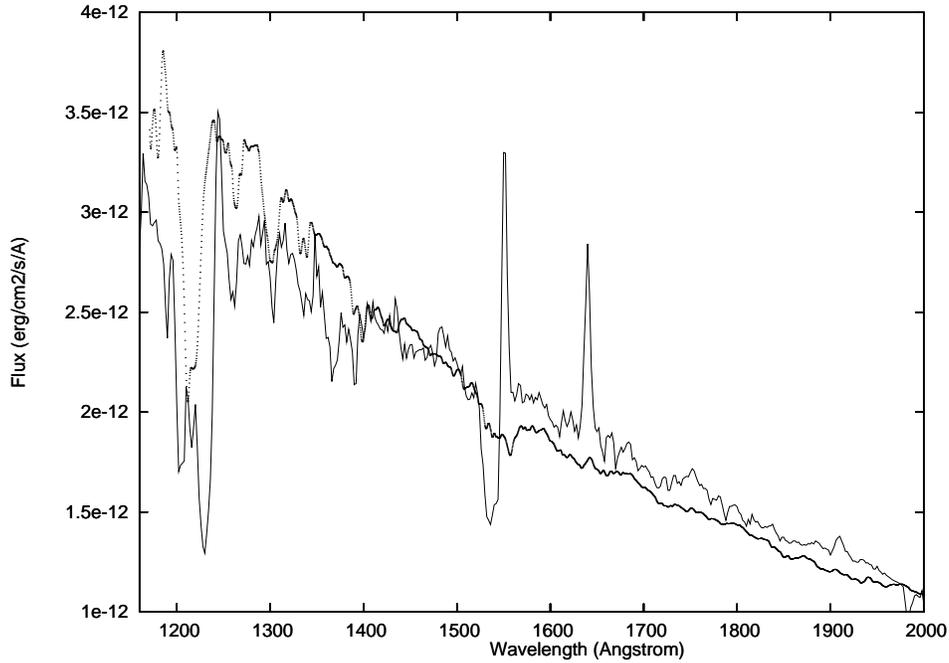}
\caption{The  reddening corrected "average"  far-UV spectrum in
1980  (continuous line) against  the "$gg$" model of  Wade \&
Hubeny (1998) for  $i$=$75.5^o$ (dots) .   The ordinate
scale of the model has been  vertically shifted  in order   to
properly fit the  data. }
\end{figure*}

\subsection {The comparison with model accretion disk spectra}

Wade \& Hubeny (1998)  have recently presented  a large grid of
computed  spectra from steady state accretion disks (AD)
in luminous CVs.
Disk spectra corresponding to twenty-six different
combinations of accretion rates  and WD masses
are computed and tabulated for six different disk
inclination angles $i$.
The wavelength coverage of the models  ranges from
$\lambda$800\AA~to $\lambda$1200
\AA, and therefore it is possible to  compare   them with  the 
IUE
spectra taken with the SWP camera ($\lambda$$\lambda$1160-1960
\AA). It should be  pointed out that some notable features in
the spectrum of HR Del, e.g. the emission lines and the P Cyg
profiles that are probably formed in regions separate from the
AD itself (chromosphere-corona, thick wind), are not present in 
the
model. Therefore, the spectral features we have  used for the
fitting are the shape of  the continuum  distribution  and the
intensity and width of the  un-displaced absorption lines (e.g.
SiII 1260 \AA ,  OI + SiII  1305 \AA ,  CII
1335 \AA ,   etc).

We have compared  most of the 156
different models with the reddening
corrected average IUE-SWP spectrum of 1979-1980 (the
spectra at the other epochs differ just in the y-scale because of
the near gray decay).
The results are only partially satisfactory:
\begin{enumerate}
\item
The observed continuum flux, when scaled to the
normalized distance of the models (100 pc) is higher than in any
model. Only model $jj$  (a massive WD  at high 
$\dot{M}$)
gives fluxes that are comparable with the observed one. 
Therefore,
as far as  the continuum is concerned, we have
limited the comparison to the slope only.
\item
No one solution  is convincingly valid  at the same
time  both for the continuum (slope)  and the depth and shape
of the absorption lines .
\item
The best solutions for the continuum (e.g. models   $ff$ with
 $i$=$75.5^o$   or  $i$=$41.4^o$,   $ j$ with $i$=$18.2^o$)
  give  line features that are too strong and  too deep.
 \item
The best  fits for the shape and depth of the lines (models  $cc$,
$hh$) indicate higher $M_{1}$  values (e.g.
$\sim$0.8$M_{\odot}$) , but give a continuum which is too hot.
\item
The best line+continuum  compromise  comes from the $bb$ and $gg$
models ($M_{1}$=0.55$M_{\odot}$) but only when seen at rather high
inclination angles., $i$=$75.5^o$ .
\end{enumerate}

Fig. 8 is a plot of THE  "average"  SWP spectrum  of 1980 together
with the "$gg$" model  ($M_{wd}$=0.55$M_{\odot}$,  from Wade and
Hubeny (1998) with $i$=$75.5^o$ .

In conclusion,  the  fits to the Wade and Hubeny models
(1998),  give  a range of acceptable values that are
"compatible" with the physical parameters of HR Del. Certainly,
models with $M_{1}$ larger than 0.8$M_{\odot}$ , (too steep
continuum, too wide lines) and models with $M_{1}$ lower than
0.55$M_{\odot}$ (too deep lines) are  ruled out. A
refinement of the grids to include  additional intermediate values
would be effective in setting tighter constrains to the
parameters. However, in any case, as we shall see below,  the
observed continuum luminosity would need to be explained by a
considerably higher accretion rate than the  10$^{-8}$ $M_{\odot}$
yr$^{-1}$ limit of the Wade \& Hubeny (1998) models.

\subsection {The $\lambda$$\lambda$ 1180-3250 \AA ~
continuum  distribution }

In a different  approach  to the study of the UV continuum  we
can assume that its distribution  can be
represented by a black-body  (BB) and/or a power-law (PL) and we
can try to determine the parameters that provide the best fit to
the UV data.

\begin{figure*}
\centering
\includegraphics{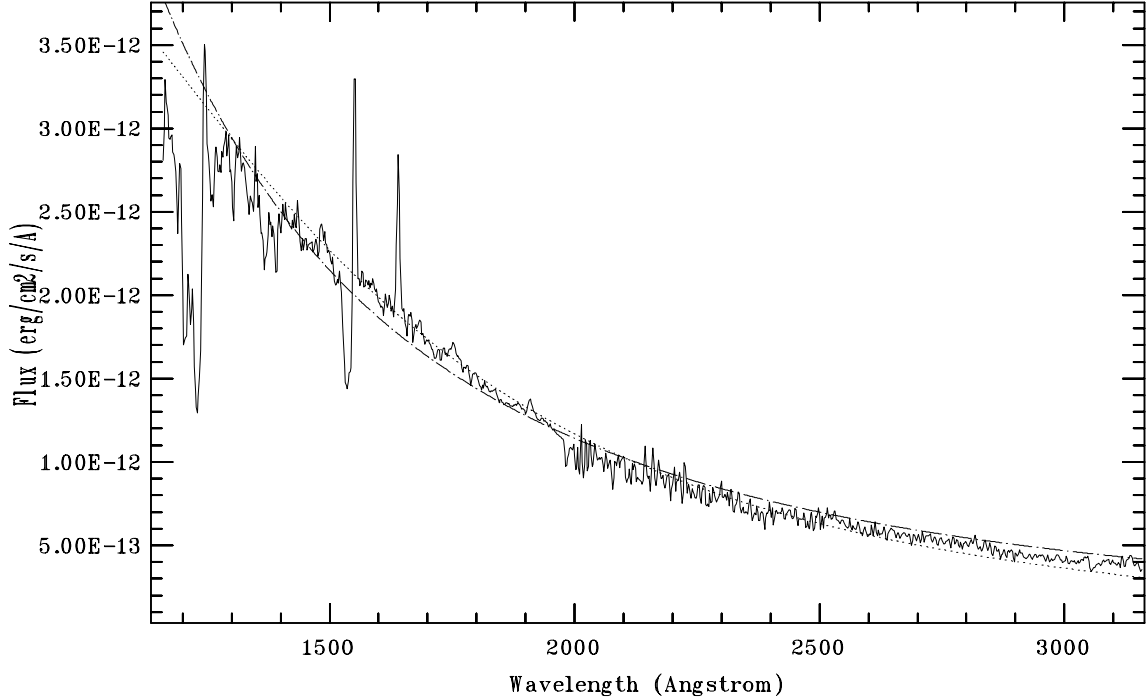}
\caption{The "average" 1980-1988  UV spectrum (see Fig.1) after
correction for E$_{B-V}$=0.16. A 33900 K black-body curve  (dots)
 and a power-law curve with index $\alpha$=-2.2   (dot-dashed
line) are superposed.}
\end{figure*}

The UV continuum  is clearly defined longwards of $\lambda$1400 \AA
~ where  only a few and well known spectral lines  are present.
For a correct  definition of the  continuum
below $\lambda$1400 \AA~ where the continuum position is
more uncertain  we have been helped by a preliminary
identification of the spectral features (most of them being
absorption lines, as mentioned in  Sect. 3).  Also, an
impersonal method (the icfit task in IRAF)  has yield a
similar result. A curve  fitting to this  continuum
has been made using several data analysis packages (gnuplot,
Dataplot, Grace) but especially  the IRAF
stsdas.analysis.fitting.nfit1d  application.
This  code fits 1-dimensional non-linear functions (BB, PL, and
combinations of them) to the data. The non-linear fitting can be
performed by any of two algorithms, either one of which can be
used to minimize chi-squared: the downhill simplex ("amoeba") or
the Levenberg-Marquardt algorithm. The fitting algorithm needs
initial guesses for the function coefficients (parameters) that
can be entered interactively.
The best single curve fit to the
(impersonal) UV continuum distribution of HR Del corresponds  to
a BB with T$\sim$33,900 K, while  a BB with a slightly lower T is
still the best fit to the "hand traced" continuum. A power-law
distribution with
$F_{\lambda}$=2.0$\times10^{-5}\times\lambda^{-2.20}$ (erg
cm$^{-2}$s$^{-1}$ \AA$^{-1}$) is also a good approximation with a
small uncertainty of $\pm$0.05  in the index, depending again on
the assumed continuum.
In any case, the  PL fitting  gives a
spectral index ($\alpha$=2.20) that is near the Lynden-Bell law
for a standard disk ($\alpha$=-2.33) as  already found by 
Friedjung
et al (1982)  from the early IUE spectra.
A composite (BB+PL) fitting is
obtained with  a BB with 35100 K and a power-law with index -1.84,
but does not show any significant improvement to the good fit
provided by the single BB at 33900 K.
It should be pointed out that there is neither  any indication nor
any
requirement for a high T component: we have tried to "force" such
a presence by making  such a guess for the initial
coefficients but the algorithm has always pointed toward lower
temperatures.

As a matter of fact, the best fit
to the continuum, the one with the lowest residuals, comes from the
combination of two BBs, one with T=34700 K and the other
with T=5200 K. The contribution of the latter starts at
wavelengths larger
than about $\lambda$2600 \AA~and its extrapolation
becomes gradually dominant toward the optical range. It is
tempting, "prima facie", to interpret this  behavior in terms of a
two component continuum with the  hot component (the AD)  that
dominates the  UV, and  a  cool component, presumably the
companion star,  that is associated with
the optical  magnitude $m_{v}$=12. However, simple calculations
show that an object with $T$=5200 K (probably a K0V star), and
$R$=0.55$R_{\odot}$  at the distance of 850 pc would  produce  a
flux of about 1.0$\times$10$^{-15}$ erg
cm$^{-2}$s$^{-1}$ \AA$^{-1}$ at $\lambda$
5480 \AA, the effective wavelength of the V filter for a low
temperature star. This flux can be converted to m$_v$ using the
zero mag absolute calibration   (Gray, 1992)

\begin{displaymath}
\log F_{\lambda}=-0.4\cdot V -8.45 ~~ ({\rm erg ~ cm^{-2}s^{-1}
\AA^{-1}}) \end{displaymath}

This yields $m_{v}$=16.0, that is, $m_{v}$=16.5 after
"reddening" it  by A$_{V}$=0.50. The same
conclusion can be directly obtained by estimating  the
apparent visual magnitude of a K0 V star ($M_{v}$=+5.9 ) if  $d$
=850 pc and $A_{v}$=0.50 mag:  the resulting value is close
to $m_{v}$=16. In order to produce  $m_{v}$$\sim$12 the alleged
cool component with $T$=5200 K  requires a radius of about 6
$R_{\odot}$, a requirement that  would imply the presence of
a very evolved secondary,   not compatible  with the well
defined orbital period  P=0$^{d}$.214165.
Therefore, we consider the requirement for  a  cool
component just as an analysis "artifact" that  might simply arise
from the  fact that  the continuum distribution curve is not a
perfect black body.

In conclusion: a BB with  $T$=33900 K , or alternatively a
power-law with $\alpha$=-2.20 represent a good fit to the
observed UV continuum distribution.  Other  fits, although
apparently more accurate, lack on  a physical basis.
We take the temperature $T$=33,900 K  as indicative
of the "mean" accretion disk  temperature.

It is  remarkable that the  extrapolations at $\lambda$5450  \AA~,
(the $\lambda_{eff}$ of the V-band for a hot star) of either
of these single function distributions that best fit the observed
UV continuum (black-body, power-law) yield optical fluxes
(4.7x10$^{-14}$ erg
cm$^{-2}$s$^{-1}$ \AA$^{-1}$  and 11.8x10$^{-14}$erg
cm$^{-2}$s$^{-1}$ \AA$^{-1}$) that after conversion to magnitudes
and proper "reddening", as described previously,  give
 $m_{v}$$\sim$12.7 and $m_{v}$$\sim$ 11.7 respectively.

Therefore,
the "quiescent" optical magnitude at 12$^{m}$ corresponds with
the "tail" of the UV continuum distribution, that is  the
tail of the "hot" source.

The optical flux (5$\times$10$^{-14}$ erg cm$^{-2}$ s$^{-1}$
\AA$^{-1}$) found by Ringwald et al. (1996) is close to
our UV extrapolated values.  The presence of optical nebular
lines and the hot continuum reported in the same study  confirm the
persistence of the hot source observed with IUE.

 Fig  9 is a plot of the  de-reddened "average" spectrum together
with the single power-law and single BB fittings.
Toward  longer wavelengths the single black body distribution
with T$\sim$33,900 K  falls slightly below the well defined
continuum, while the power-law falls above it.

\subsection {$L_{UV}$, $M_{v}$ and $\dot{M}$ }

The reddening  corrected UV-integrated ($\lambda$$\lambda$
1160-3250 \AA) continuum flux is of
2.5$\times$ 10$^{-9}$ erg cm$^{-2}$s$^{-1}$.
If $d$=850 pc then   the reddening corrected total  UV luminosity
is  2.2$\times$10$^{35}$erg s$^{-1}$  that is,  $L_{UV}$$\sim$
56$L_{\odot}$.

We  recall that in the best observed old novae   $L_{UV}$   is
in the range 1-20 $L_{\odot}$ (Gilmozzi et al. 1994; Selvelli et
al. 2003)  the highest values being associated with systems seen
near pole-on,  while for systems at intermediate inclination, like
HR Del, a typical value is $\sim$5$L_{\odot}$.   See also
Friedjung and Selvelli (2001)  for  other  considerations  on the
peculiar luminosity of HR Del.
Therefore,    HR Del   appears  as  the brightest member  in the
UV  among ex novae  and  is challenged only by the two recurrent
 novae T Cr B   and T Pyx ,  at  $L_{UV}$$\sim$50$L_{\odot}$
   (Selvelli  et al. 1992) and  $L_{UV}$$\sim$98$L_{\odot}$
(Gilmozzi et al. 2003),   respectively.
 Incidentally, we point out that
the overall (line and continuum) UV spectral appearance of HR Del
would be quite similar to that of the recurrent nova T Pyx, were it
not for the absence in the latter star of the  P Cyg absorption
components in the NV and CIV resonance lines  (See also Chapt.
6.7).

 In the following we confidently assume  that the observed UV
luminosity comes from  an  accretion disk  heated  by   viscous
dissipation of gravitational  energy.   This
sounds reasonable in view of the general behavior in other similar
objects, old novae and nova-like stars, and  is supported by the
fact that the UV continuum distribution is close to  that expected
for a "standard" optically thick AD, as well as that predicted by
an early simplistic model of an irradiated AD (Friedjung 1985).  In
Chapt. 6.6 we consider also some problems that arise if we assume
the  presence of a hot, bloated WD as
the main source for the UV continuum.

From $m_{v}$=12, assuming
$d$=850 pc,  and A$_{V}$=0.50  ($E_{B-V}$ =0.16) we obtain
$M_{v}$=+1.85  for the "apparent" absolute magnitude.  Supposing
that  the visual radiation  comes from a non-irradiated disk
we   correct for the inclination (i $\sim$40$^{o}$)   by a
factor  $\Delta {M_{v}}$=0.45,  an average from  the Paczynski and
Schwarzenberg-Czerny (1980),  Warner
(1986) and Webbink et al. (1987) relations,  to obtain an
absolute magnitude avertaged over all directions of
$M_{v}$=+2.30.  This  V is  brighter by about 1.7 magnitudes
than the mean absolute magnitude  of nova remnants in the
same speed class,  whose average value is centered on
$M_{v}$=+4.0 (See  Fig.  2.20 and  Table 4.6  of  Warner 1995).
This V is also brighter  by about 2.4 magnitudes than the mean
absolute magnitude of novae at minimum, as  obtained  from the
values given    in Table 6 of Downes and Duerbeck  (2000).
In this list,  only
one object  (BT Mon)  appears brighter than HR Del, but,   being a
high inclination system,  its V value  could be affected  by a
substantial uncertainty in the correction.

The   "pole-on"   ($i$=0)  absolute visual  magnitude   of
HR Del  obtained by the  same  correction
method (but with a negative value   for a  disk viewed
"pole-on'')  is close to  $M_{v}$=+1.35.
This value is also brighter  by about  1.3 magnitudes  than
the average value for nova remnants seen   "pole-on"  (See Fig 2.20
 of  Warner 1995).
  
Therefore, HR Del at the "quiescent"  $m_{v}$=12 is
one of the   brightest nova remnants also in the optical range.
This result   is not very surprising if one considers
the peculiar UV  luminosity of HR Del and the fact that, as
mentioned before, $m_{v}$ comes from the same source. 

We emphasize that the  corrections for inclination mentioned
above are  valid only if the observed continuum comes from a
disk that
is heated by viscous dissipation of gravitational
energy.  The
luminosity of a bare WD clearly needs no
inclination
correction. Lapidus and Sunyaev (1985) discuss the case
of an
irradiated disk, giving an inclination correction, which
leads to
the increase of the calculated bolometric luminosity of
the
irradiating   central object by a factor near 3.6.

We recall that Warner (1987) has given $M_{v}$ $\sim$+3.9  (and
also  + 4.2) for HR Del as  an ex-nova , but he has assumed
$d$=285 pc,  a value that seems difficult to reconcile with all
other estimates.

In principle, the mass accretion rate  $\dot{M}$  in a viscously
heated disk  can be  estimated    from a comparison
between the observed spectral distribution and that of proper     
models. However, the number of parameters in any disk model is
quite large and the various methods of fitting the data to the
models do not generally provide unequivocal results.
On the other hand, if the total disk luminosity and the mass of the
WD are known, $\dot{M}$ can be directly calculated  from the
relation

\begin{displaymath}\dot{M}=\frac{2~ R_1~ L_{disk}}{G~
M_1} \end{displaymath}

In this case,  the estimate of $\dot{M}$  is not model dependent
but requires the knowledge of the bolometric accretion luminosity
$L_{disk}$ and of $M_1$.    The observed $L_{UV}$ can
provide a first estimate of $L_{disk}$,   $L_{UV}$,  being
a lower limit to the bolometric disk
luminosity $L_{disk}$.

We recall that  most of the accretion
luminosity is emitted in the IUE-UV  range: radiation at
wavelengths shorter of Ly${\alpha}$ is strongly absorbed and the
energy is redistributed toward longer wavelengths (Nofar et al.
 1992).
If the mean accretion  disk temperature is close to 33900 K,  with
the assumption  that the distribution is close to that of a BB,
one can easily  see that the peak of the distribution  falls near
$\lambda$ 850  \AA~  and that  about
half of the total power is emitted shortwards of about
$\lambda$1200 \AA. It seems therefore
justified to assume $L_{disk}$$\sim$2$L_{UV}$,  that is ,
$L_{disk}$$\sim$112L$_{\odot}$. Also, a detailed examination  of
the Wade \& Hubeny (1998) grid of  model spectra for
the relevant cases shows that  approximately only one half of the
total continuum energy is emitted shortwards of
$\lambda$1160\AA.

If we adopt  $L_{disk}$$\sim$112$L_{\odot}$,
together with   $M_{1}$$\sim$0.65
$M_{\odot}$ and $R_{1}$$\sim$0.0125$R_{\odot}$ we obtain
that $\dot{M}$$\sim$1.4 $\times$$10^{-7}$$M_{\odot}$
yr$^{-1}$.

An independent determination of the mass accretion rate can be
obtained
through the  $\dot{M}$-HeII 1640 \AA~luminosity relation given in
Table 2 of  Patterson \& Raymond  (1985). The average
(de-reddened) flux on earth in the HeII 1640 line is of
$7.0\times 10^{-12}$ erg cm$^{-2}$ s$^{-1}$   and the
corresponding luminosity is 6.1$\times$10$^{32}$ erg
s$^{-1}$. If   $M_1$=0.7$M_{\odot}$ this corresponds  to a
mass accretion rate $\dot{M}$ $\sim$10$^{19}$ gr s$^{-1}$,  that
is $\dot{M}$$\sim$1.4$\times 10^{-7}$ $M_{\odot}$
yr$^{-1}$, in excellent (if not surprising)
agreement with the estimate based on the UV continuum.

The mass accretion  rate $\dot{M}$ can be also estimated  from the
"average" $M_{v}$.  using the
relation of  Webbink et al. (1987):
\begin{displaymath}\dot{M}=\frac{10^{0.6(-9.48-M_v)}}{M_1}
\end{displaymath}
Substitution of the SPECIFIC  values for HR Del  ($M_{1}$=
0.65$M_{1}$,  $M_{v}$=+2.30 )    gives
$\dot{M}$=1.31$\times$10$^{-7}$ $M_{\odot}$ yr$^{-1}$.   We point
out that the   $\dot{M}$ obtained by this  method   comes from the
adoption of a   $M_{v}$ value   that corresponds to an
 "average"
disk  ($i$$\sim$ 58$^{o}$).

The   remarkable  fact  that  three  different methods
have yield very similar $\dot{M}$  values  speaks in
support  of the  reliability of the results  (and of the methods).

An estimate of the  mean size of the
emitting surface can be obtained using the "mean"  disk
temperature $T$$\sim$33900 $K$ (see Sect. 6.3)  and  the
total disk luminosity  $L_{disk}$$\sim$2$\times$$L_{UV}$
$\sim$112$L_{\odot}$.  We obtain $R$
$\sim$0.36$R_{\odot}$,  in  good agreement with the
expected size of  an accretion disk  in a  cataclysmic variable.

It must be stressed   that all the  quite  high  $\dot{M}$
values obtained above come from  the adoption of a
rather conservative  (low)  value for the distance.
In order to  obtain values  for  $L_{UV}$,  $M_{v}$ and $\dot{M}$
close to those found  in other ex-novae  one should   adopt   a
much lower value for the distance  ($d$$\le$ 400 pc),  but this  is
sharp contrast with all more recent estimates that give  d$\sim$1
kpc.

Using the values for $\dot{M}$, $M_1$, and $R_1$ derived
previously, we can estimate the value of the "maximum" disk
temperature from the common relation
\begin{displaymath}
T_{disk}^{max}=0.488  \left(\frac{3 GM_1
\dot{M}} {8 \pi \sigma  R{_1}^3}\right)^{1/4}
\end{displaymath}
We obtain $T_{disk}^{max}$$\sim$1.08$\times$10$^{5}$ K,  to be
compared with the maximum disk temperature in the  relevant
models of  Wade  and Hubeny (1998)  that ranges from   $T$ $\sim$
 39,110 K  (model $jj$)   to $T$ $\sim$  69,560 K  (model $hh$)
and with the "mean" disk temperature $T$=33900 K estimated in the
previous chapter.

\subsection {The pre-nova  and the ex-nova }

The comparison between  the pre-nova
and the post-nova magnitudes has been the focus of various
investigations  that have led to non-unequivocal results and
interpretations.

Payne-Gaposchkin (1957) from a study of 12 novae concluded that the
pre-nova and post-nova brightnesses (years after outburst) are
more or less  the same. Robinson (1975), from a slightly larger
sample,  fundamentally confirmed that the mean magnitudes of the
pre-nova and the post-nova are essentially the same, but  also
reported for many novae a moderate rise in brightness in the 1-15
years prior to the eruption.
Shara et al. (1986), in the context of the "hibernation" theory ,
claimed that  pre novae become "bright" (at  $M_{v}$ near + 4.0,
the average  absolute magnitude of novae before and after eruption)
a few decades before the eruption, and old novae remain bright at
near the same $M_{v}$ for many decades after outburst. The  high
values observed in CN before and after eruption are allegedly
produced by the resumption of contact of the red dwarf with its
Roche lobe (after the drain of angular momentum from the system by
gravitational radiation  or by magnetic braking by a stellar wind)
and by the  irradiation of the secondary, respectively. The
hibernation theory predicts an order of magnitude decrease in the
quiescent luminosity over a long timescale   (several decades or a
century) (Kovetz et al. 1988) and therefore  good
data on the secular brightness evolution of post-novae  are
required to test the theory. Vogt (1990) in a study of 97 well
observed galactic novae found that during the first  130 years
after the eruption galactic novae show a slow decrease in
brightness with a rate  of 2.1 mag per century  and interpreted
this as  evidence in support of hibernation.
Duerbeck (1992), from a detailed study  of a
limited number of  well selected post-novae, based on
observations obtained at least 20 yr after outburst, found instead
a decline rate of 10.4 mmag/year and a behavior compatible with
hibernation

The situation seems  however controversial:  Weight et al. (1994)
examining near IR photometry for a number of ex-novae  found no
correlation between $\dot{M}$ and time since outburst.
Recently, Duerbeck (1995) has reviewed the general
assumption that m(pre-nova)=m(post-nova)   and found
that most novae, some years after outburst are about  2$^{m}$
brighter than before outburst, and  decline gradually  to
pre-outburst brightness in about 10-30 years of exponential decay.

In the case of HR Del, we recall that Stephenson (1967) found
$m_{v}$ about  12 for the pre-nova magnitude  and that
Wenzel (1967), from the inspection of Sonnenberg plates  taken
between 1928 and 1966  found the pre-nova to be only slightly
variable around a mean photographic magnitude of 11.9. Van den
Bergh \& Racine (1967) also found evidence of small variations in
brightness from a comparison between PSS plates obtained in 1951
and 1953 and pointed out that the pre-nova was very blue.

HR Del is one of the few novae for which pre-nova
spectra are available:~ Stephenson (1967) classified the pre-nova
 continuum (seven years prior to outburst) to be that of an O or
early B star while Hutchings (1968) estimated the
temperature of the pre-nova star to be  $\sim 32000$ K.  Barnes \&
Evans (1970) from  photographic photometry  on Palomar Sky Survey
plates found (B-V)=-0$^{m}$.18 for the pre-nova in 1951.  After
correction for E$_{B-V}$=0.16,  this gives   (B-V)$_o$=-0.34  in
good agreement with the value  expected from an object with T
$\sim$ 33,000 K.  Also, Seitter (1990), from an uncalibrated
low-resolution pre-nova spectrum of HR Del found evidence of
high  temperature  continuum.

After the  outburst  of  1967 the  extra
slow decline in $m_{v}$ of HR Del lasted for about 15 years with an
asymptotic approach toward  the pre-outburst  value
($m_{v}$=12) that was reached around 1981-1982.
Thus,  most IUE observations were taken after the return of
the nova to the pre-outburst  V-magnitude.

In conclusion, for many years  \emph {before}
the 1967 outburst HR Del has been  at the \emph {same}  $m_{v}$ and
$T$ values ($m_{v}$$\sim$12.0 and $T$=33,000 K) as during the
post-nova stages corresponding to most  IUE observations.

Recalling (Sect. 6.3)  that the "quiescent"
magnitude at $m_{v}$=12 comes  from the "tail" (extrapolation to
$\lambda$ 5450 \AA) of the observed  hot  UV continuum,
which is close to both a   power-law  and  a 32900 K  black-body,
(and is not due to  the cool companion which  can account for only
a small fraction of the required V flux) it follows that  the star
during the pre-nova stage at 12$^{m}$  had necessarily  the high
$T$, $L_{UV}$, $M_{v}$ and   deduced $\dot{M}$ values found in the
present study.

We point out these peculiar aspects in the behavior  of HR Del:
 \begin{enumerate}
\item
 As a post-nova,  HR Del  at   $L_{UV}$=56
L$_\odot$   and  $M_{v}$=2.30  is   brighter than its colleagues
both in the
optical and in the UV. One could
expect this behavior if the post nova had
remained  in an "excited state" at, say, $m_{v}$=11.6,
because of the extreme slowness of the outburst phases and of a
corresponding extra slow decline toward the pre-outburst
brightness.
Instead,  this "bright" state corresponds to  the return to
the  pre-outburst magnitude.
\item
 As demonstrated previously, during at least 70 years before
the 1967 outburst  at $m_{v}$ $\sim$12,  HR Del was necessarily
at the same high T, $L_{UV}$,  and deduced  $\dot{M}$,  as 
derived
from
the IUE and other  recent observations . This sets a time
constraint to a possible pre-outburst onset of  high $\dot{M}$.
\item
Apparently,  there is no indication of any
  further decline in  optical brightness during the last 18
years,  that is, after the return to $m_{v}$$\sim$12.0.
\end{enumerate}
We cannot but  point out that, in the context of the
"hibernation" theory, it is difficult to understand how two
different processes (the contact induced $\dot{M}$ in pre-outburst
and the irradiation  induced $\dot{M}$ in the post-outburst) could
have led  to the \emph{same} $\dot{M}$
values,  as to produce the \emph{same} pre-and post-outburst
magnitudes. The process of irradiation of
the secondary should have been especially important for  HR Del
because of the extra-slow character of the outburst that has kept
both the WD and the companion exposed for a long time to the hard
radiation of the nova.

In other words it is not clear how, many years before outburst
HR Del  managed to  be at near the same
high  values for $\dot{M}$, L, T  as in the post-nova
stage.

It is worth recalling  that  the  magnitude range
$A$=$m_{v}$(outburst)-$m_{v}$(post-nova)  of HR Del  ($\sim$8 mag)
is
within the average behavior (Warner, 1995)  of other members of the
extra slow class . This parameter however depends  on the
difference between the pre-outburst and the post-outburst values,
and not on the absolute values.

\subsection  {A still burning  white dwarf ?}

Hydrodynamic models of nova outburst predict that, after the
onset of the outburst, hydrogen nuclear burning  should continue on
the surface of the WD for as long as  100 years, with a strong
inverse dependence on the mass of the white dwarf.
Real novae instead seem  quite inpatient and are able  to
shorten considerably this phase through some not well established
mechanism (enhanced mass-loss via radiation pressure-driven  winds,
common envelope ejection, magnetic  fields,  etc).
Combined UV + soft x-ray observations (Gonzalez-Riestra et al.
1998, Vanlandingham et al. 2001) have  indeed shown that most novae
decline in bolometric luminosity within a few years from the
outburst.

Sion \& Starrfield (1994) have presented  theoretical results on
the processes  of   compressional heating,   low-level
hydrogen burning,  but not accretion   on very hot, low-mass WD,
and found that these processes power the models.   The
luminosities are  a few times 10$^{37}$ erg s$^{-1}$ at certain
epochs, being much less ( e.g.   $\sim$10$^2$ erg s$^-1$)
at other  times. They suggest that these models could be relevant
to the class of ultrasoft X-ray sources and related objects.

In this framework, it is  conceivable that some aspects in
the observed behavior of HR Del, e.g. the observed high UV
luminosity  and the strong P Cyg profiles, indicative  of
outflow, could be  interpreted as evidence that
 continuous hydrogen burning is taking place on the
surface  envelope of a low mass  white dwarf.   Also,  the
conservative  estimate for the
absolute visual magnitude of  HR Del
in quiescence ($M_{v}$=+2.30, Section 6.4)   would
place it in the region of  the supersoft binaries  according to
Fig.2 of Patterson (1998).

However, both the slope of the continuum (that is indicative of  a
temperature  of the order of  33900  K)  and the character of the
line spectrum (especially the relative intensities of the CIV
 1550  \AA~ and HeII 1640 \AA~emission lines, and the
weakness of  the NV  1240 \AA~line) are hardly compatible with the
presence of a very hot WD with T$\sim$2$\times10^{5}$ $K$, as
expected in this case,  unless some kind of reprocessing of the
"hard" radiation takes place, e.g.  in the  irradiation of the
accretion disk.  The presence of  a compact and very hot
continuum source, whose contribution toward the UV-optical regions
is relatively small, could  also be invoked but in any case the
absence or weakness of emission lines of high excitation is
disturbing. Also, the observed absorption lines of SiII at 1260
\AA,  and CII at 1335 \AA, etc., are indicative of stellar spectra
about B2- B7 with  T $>$20,000 K and T $<$ 32000 K,   in fair
agreement with the  T from the continuum, and indicative of a
multi-T structure, as expected in an AD.

Another argument against the hypothesis of a still burning WD
lies in the difficulty of  reconciling  the hypothesis
of an enduring phase of continuous burning on the WD surface with
the fact that  the system  (ex-nova) is exactly  at the \emph
{same} pre-nova optical magnitude.  This would imply   that the
pre-outburst "quiescent" magnitude at $m_{v}$$\sim$12  (almost
constant  in accordance with  all "historical" data available)
(line and continuum) a phase of stable burning  which might even
have endured since the previous outburst.

\subsection {The wind features and their origin}

The absorption
component in the CIV 1550 \AA~ doublet reaches
 $v_{edge}$$\sim$5000 km $s^{-1}$ and its shape suggests the
presence of two
structured components  displaying their maximum depth near the line
center. A similar behavior is displayed by the NV
1240  \AA~ doublet, but because of the proximity of Ly$_{\alpha}$
a precise measurement is difficult. It should be noted that this
high $v_{edge}$ value contrasts with the moderate outflow
velocities recorded at the time of the outburst when  most
spectroscopic observations showed velocities in the range from -200
to -700 km s$^{-1}$ (Hutchings 1968).
In any case the presence for at least 25 years after outburst of
strong wind features in the line spectrum of  HR Del is surprising.
In  the first  IUE spectra  of 1979 they  were
interpreted in the framework  of a continuing mass outflow in the
outburst of a very slow  nova (Hutchings 1979).  However, the
absorption profiles were still present, although slightly weaker,
in the last IUE spectra of 1992 and presumably they still are.
\begin{figure*}
\centering
\includegraphics{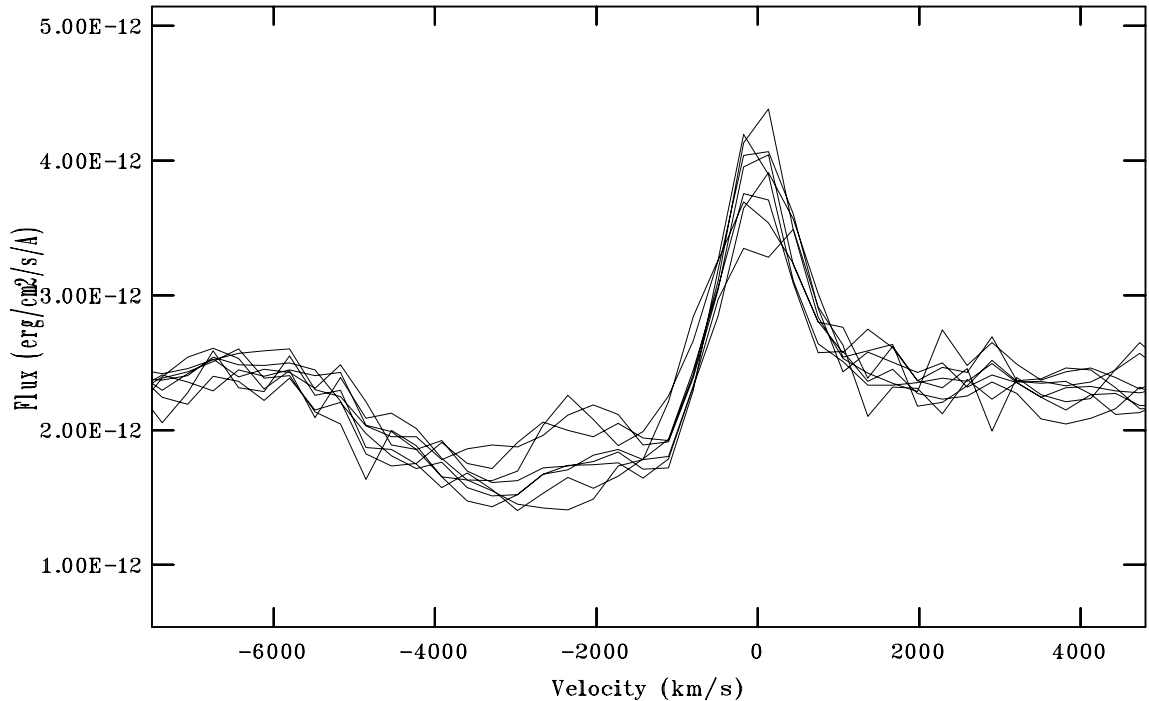}
\caption{The short-timescale changes in the wind absorption
features of the CIV 1550 \AA~ resonance doublet. The spectra belong
to the sequence of  21 Aug.,1980 and the time separation between
two successive SWP exposures is about 49 minutes.}
\end{figure*}

The phenomenology of the short-time variations of the wind
(as judged from the behavior of the  CIV and NV absorption
components) indicates the presence of an inhomogeneous (structured)
outflow   with  irregular variations on short timescale and
the sudden ejection of puffs of optically thick material.
This behavior is
similar to that observed in other CVs and especially in V603 Aql.
In this context, we recall that Friedjung et al.
 (1997), from a GHRS study of V603 Aql, suggested that the
emission and absorption components of the P Cyg profiles originated
from two separate physical regions, i.e.  a chromosphere corona
which surrounds the disk, and a conical-shaped wind region,
nearly perpendicular to the disk itself.
In a follow-up  IUE study on the variations in the wind features of
V603 Aql,  Selvelli et al. (1998) pointed out  that the
outflow  should take place through the sudden ejection of puffs of
optically thick material, with timescales  on the order of tens of
minutes or less. No definite periodicities were associated with the
presence of the wind nor any correlation between the  absorption
features and the UV flux modulation detected. A similar
conclusion was reached by Prinja et al. (2000) from  GHRS data
of V 603 Aql taken with high time resolution: they observed
variability on timescales of minutes and suggested an empirical
picture of stochastically variables structures in the outflow with
no evidence for any cyclic modulation in the absorption components
properties.

 The similarity in the class of  wind phenomena  between
 HR Del and V 603 Aql  suggest that the same basic mechanism,  that
is, a collimated outflow (conical wind) from the inner accretion
disk region  is present in both stars. The absence of any
correlation between the absorption components  and the changes in
the UV  flux  or the EUV flux (from the HeII  1640  \AA~line)
should set some definite constrains on the physical interpretation.
Prinja et al. (2000) concluded that the  irregular absorption
episodes of V603 Aql defied a clear physical interpretation, and a
similar conclusion, regrettably,  seems  valid for HR Del also:
the origin and region of formation of the wind features are not
clear.
We recall that the presence of  short time
variations in the profiles of the resonance lines  has been
reported also for several other CVs systems not belonging to the CN
subclass.

In the framework of  the hypothesis of a bi-conical outflow, after
correction for the inclination ($i$$\sim$40$^{o}$), a quite high
value for the intrinsic v$_{edge}$, on the order of 7000 km
s$^{-1}$,  is obtained.  We recall that Hutchings (1980) already
noted that in HR Del the terminal velocity was higher that in
other stars except perhaps some WR stars. We recall also that a
very low system inclination is not compatible with the other system
parameters and that a \emph{lower limit} for i is about 35$^{o}$.
In other CVs, the observed terminal v is generally comparable to
the  escape velocity from the primary.  For HR Del, such high value
would indicate a massive WD, but the outburst behavior and data on
the system suggest that the WD in HR Del is rather
undermassive compared with other CN ($M_{wd}$$\sim$0.65
$M_{\odot}$)   and a lower v$_{esc}$ is expected.  It is worth to
point out that in V603 Aql, where, allegedly, a  more massive  WD
is present, the observed $v_{edge}$ reaches a value on the order
of -2700 km s$^{-1}$, (that is,  $v_{edge}$$\sim$-2850 km
s$^{-1}$,  after correction for the low i).

The problem of an exceeding high value for v$_{edge}$ could be
alleviated only if  the outflow  took place in a spherical
geometry. However, an outflow from the whole WD is difficult to
envisage without invoking  some  kind of persistence of  TNR
activity  on the surface of a bloated white dwarf.
 We have above mentioned some difficulties connected with  this
hypothesis, at least if it is supposed that no accretion disk is
present.  In any case,  the apparent similarity in the phenomena of
the short timescale  variations in HR Del with those in V 603 and
other CVs   may favor  an interpretion in terms of phenomena
taking place in a bi-conical AD geometry.

It seems quite obvious to associate the presence of the strong
 P Cyg profiles with that of a strong UV luminosity.
In this case the high velocity  P Cyg  absorption components could
be due to a wind driven by a radiation force  in the lines.
In any case, it seems to us that the radiation force alone cannot
account for the observed wind properties of  HR Del:
the  recurrent nova  T Pyx   has nearly the same UV continuum
shape and an higher  UV luminosity than
HR Del   but shows no evidence of any absorption component
despite the fact of being observed  at a  more favorable (lower)
inclination angle ($i$$\sim$17$^{o}$).
A non-radiative factor such as  the presence of a strong magnetic
field could be invoked (see Proga 2000, Hartley et al. 2002) but no
evidence for it has been found so far in HR Del.
In addition, the presence of puffs and structures in the wind
remains a challenge to our understanding of this class of
phenomena.

Downes \& Duerbeck (2000), in
 measuring  the distance of HR Del  with the method of
the expansion parallax, have  found some evidence of
acceleration in the shell. They have interpreted this result as an
indication of the presence of a fast wind from the stellar remnant,
as supported by the P Cyg profiles detected in the first IUE
spectra (Krautter et al. 1981). On the basis of  our  previous
considerations it may seem clear that the wind geometry is not
spherical, and that if the outflow axis is perpendicular to the
disk, acceleration is expected only in the polar caps. In turn,
this would enhance the prolateness of the ejecta and this could
provide
an alternative interpretation for  the prolateness  of the
remnant found by Slavin et al. (1994) and attributed
by them to the common envelope phase during the nova outburst.
In this respect see also O'Brien et al. (2002).

In an alternative  view with respect to the origin of the wind,
we recall that Hachisu et al. (1996)
have  found that the  white dwarf begins to blow
(bi-conical) optically thick winds when the mass accretion rate
$\dot{M}$ exceeds the critical rate
 ($\sim$1.9$\times$10$^{-7}$ $M_{\odot}$yr$^{-1}$
for a 0.7 $M_{\odot}$ WD,
see also Fig. 2 in  Hachisu \& Kato  2001) at which steady
shell-burning can process the accreted hydrogen into helium.
At such rates the WD envelope is supposed to expand to R$_{ph}$
$\sim$0.1R$_{\odot}$ while T$_{ph}$ decreases
below 3.2$\times$10$^5$ K. If $\dot{M}$ decreases below
the critical value,
optically thick winds stop. If the mass transfer rate further
decreases below about 0.5$M_{crit}$, H shell burning becomes
unstable to trigger weak shell flashes.
We wonder whether the strong wind observed in HR
Del can  somehow  be associated with  this model: on the one hand
the model luminosities and temperatures are exceedingly high as
compared with those found in the present study, on the other hand
the critical accretion rate from Hachisu et al. (1996) is very
close to the $\dot{M}$ value derived  in Section 6.4.

\section {Conclusions}

There are aspects in the UV and optical
behavior of HR Del that are intriguing and makes it peculiar among
ex-novae:
\begin{enumerate}
\item
In spite of  of a  conservatively low value assumed for the
distance, HR Del appears to be  one of the brightest
 remnants (if not the brightest one)
among classical novae both in the  UV,
($L_{UV}$ $\geq$  56 $L_{\odot}$),  and in the optical,
($M_{v}$=+2.30),  assuming that radiation comes from a
non-irradiated accretion disk.  This peculiar  brightness could
be attributed to a  form of  persisting post-outburst activity
(low level thermonuclear burning ?)  associated with the extra-slow
character of the nova, were it not that most IUE  observations were
made near or after its return, around 1981, to the "pre-nova"
"quiescent" magnitude at $m_{v}$$\sim$12.
 Over the 13 years of the IUE
observations the UV continuum has declined by a factor less than
1.2.  However, the much larger decrease of the fluxes in  the
emission lines of highly ionized atoms suggests cooling of
the source of ionizing radiation, in the framework of
photoionization.
\item
 Since the optical
"quiescent" magnitude at 12$^{m}$ has its origin in the hot
component (the "tail" of the UV continuum) and not from the
late-type companion star. it follows that the pre-nova, observed at
$m_{v}$=12 for at least 70 years prior to outburst,  was at the
same (high) $T$, $L$, and $\dot{M}$ values as directly derived for
the ex-nova in the present study. It is not clear  why and how  the
pre-nova was so bright, and  we wonder whether  both the pre-nova
and ex-nova minima at $m_{v}$=12 are real  "minima" or correspond
to some  kind of   long time-scale nuclear burning as predicted by
the low WD mass models of Sion  and Starrfield (1994).
\item
The presence  of strong P Cyg profiles in the CIV and NV resonance
lines  with  an observed  $v_{edge}$ $\sim$ -5000 km s$^{-1}$   is
remarkable. This behavior is intriguing  because HR Del is the
only ex-nova that  permanently displays such strong
features and because in other CVs the P Cyg profiles give
lower velocities and are generally observed in low inclination
systems only.
After correction for the inclination $i\sim 40^{o}$ the situation
becomes worse and a  disturbingly high v$_{out}$ $\sim$-7000 km
$s^{-1}$ is obtained, a value that is in sharp contrast with the
relatively low mass of the white dwarf.
\end{enumerate}

In conclusion: on the basis of the observed behavior of HR Del we
have in this paper favored an interpretation in terms of
phenomena that are powered by accretion. However, we cannot
rule out an alternative interpretation based on phenomena that
take place on a still burning white dwarf  (surrounded nevertheless
by an accretion disk) which also burns at the same very low rate
before a major outburst and perhaps all the time between major
outbursts.

\begin{acknowledgements}
We gratefully thank Ed Sion for a fruitful discussion during the
final stage of this article.
\end{acknowledgements}

\end{document}